\documentclass[pdflatex,sn-mathphys-num]{sn-jnl}
\usepackage{graphicx}%
\usepackage{multirow}%
\usepackage{amsmath}
\usepackage{amssymb}
\usepackage{amsfonts}
\usepackage{amsthm}%
\usepackage{mathrsfs}%
\usepackage[title]{appendix}%
\usepackage{xcolor}%
\usepackage{textcomp}%
\usepackage{manyfoot}%
\usepackage{booktabs}%
\usepackage{algorithm}%
\usepackage{algorithmicx}%
\usepackage{algpseudocode}%
\usepackage{listings}%
\usepackage{orcidlink}
\usepackage{datetime2}%

\begin{document}

\title[Static Magnetic Properties of 
Cryogel\textsuperscript{{\tiny{\textregistered}}} 
and Pyrogel\textsuperscript{{\tiny{\textregistered}}}at 
Low~Temperatures and in High~Magnetic Fields]{Static Magnetic Properties 
of Cryogel\textsuperscript{{\tiny{\textregistered}}} 
and Pyrogel\textsuperscript{{\tiny{\textregistered}}} 
at Low~Temperatures and in High~Magnetic Fields}

\author[1,oa,x]{\fnm{Caeli L.} \sur{Benyacko}}\email{benyacko@ucsb.edu}
\author[1,y]{\fnm{Garrett T.} \sur{Hauser}}\email{Garrett.Hauser@colorado.edu}
\author[1]{\fnm{Raven J.} \sur{Rawson}}\email{rr.jewel00@gmail.com}
\author[1]{\fnm{Alan J.} \sur{Sherman}}\email{alanshermancas@gmail.com}
\author[1]{\fnm{Quinton L.} \sur{Wiebe}}\email{quinton@qleeholdings.com}
\author[3]{\fnm{Krittin} \sur{Poottafai}}\email{krittinpoottafai@ufl.edu}
\author*[3,ob]{\fnm{Daniel R.} \sur{Talham}}\email{talham@chem.ufl}
\author*[1,2,4,0c]{\fnm{Mark W.} \sur{Meisel}}\email{meisel@phys.ufl.edu}

\affil[1]{\orgdiv{Department of Physics}, \orgname{University of Florida},
\orgaddress{\street{2001 Museum Road}, \city{Gainesville}, \state{FL} 
\postcode{32611-8440}, \country{USA}}}
\affil[2]{\orgdiv{MagLab High B/T Facility}, \orgname{University of Florida}, 
\orgaddress{\street{2001 Museum Road}, \city{Gainesville}, \state{FL} 
\postcode{32611-8440}, \country{USA}}}
\affil[3]{\orgdiv{Department of Chemistry}, \orgname{University of Florida}, 
\orgaddress{\street{165 Buckman Drive}, \city{Gainesville},  \state{FL}  
\postcode{32611-7200}, \country{USA}}}
\affil[4]{\orgdiv{Institute of Physics, Faculty of Science}, 
\orgname{P.~J.~\v{S}af\'{a}rik University},
\orgaddress{\street{Park~Angelinum 9}, \postcode{040~01} 
\city{Ko\v{s}ice}, \country{Slovakia}}}
\affil[oa]{ORCiD \orcidlink{0009-0009-6878-620X}{0009-0009-6878-620X}{}}
\affil[ob]{ORCiD \orcidlink{0000-0003-1783-5285}{0000-0003-1783-5285}{}}
\affil[oc]{ORCiD \orcidlink{0000-0003-4980-5427}{0000-0003-4980-5427}{}}
\affil[x]{Present address: Materials Department, University of 
California, Santa Barbara, CA 93106-5050, USA}
\affil[y]{Present address: Department of Applied Mathematics, 
University of Colorado, Boulder, CO 80309-0526, USA}

\abstract{The static magnetic properties of the silica-based aergoels of 
Cryogel\textsuperscript{\textregistered} and 
Pyrogel\textsuperscript{\textregistered}, 
manufactured by Aspen Aerogels\textsuperscript{\textregistered}, 
were measured over a range of 
temperatures (2~K $\leq$ {\emph T} $\leq$ 400 K) and in 
magnetic fields up to 70~kG.  These data and a model of 
the responses are reported so these properties are familiar 
to others who may benefit from knowing them  
before the materials are employed in potential applications.
}

\maketitle

Version of~ \today ~at~ \DTMcurrenttime

\section{Motivation}\label{sec1}

Using the phrase ``low temperature aerogels'' 
during a search for thermal isolation materials, 
a paper by a CERN-based research team appeared 
\cite{Ilardi_2020}, which reported studies 
of the thermal conductivity of 
Cryogel\textsuperscript{{\tiny{\textregistered}}},
but the magnetic properties were not found in any database.  
While learning more about the low-temperature 
insulation trademarked as 
Cryogel\textsuperscript{{\tiny{\textregistered}}} 
 by Aspen Aerogels \cite{Aspen}, the high-temperature counterpart 
Pyrogel\textsuperscript{{\tiny{\textregistered}}} was 
identified as potential insulation for a 
materials-processing in high magnetic-fields station
that was being constructed \cite{Flynn2022,Flynn2025}.  
Consequently, a sample-pack was purchased, and undergraduate research 
students were trained to acquire, analyze, and report the magnetic data 
\cite{Sherman2022,Wiebe2022,Benyacko2025}.  

In an attempt to be useful to others who may be interested in the low 
temperature and high magnetic field results, 
this brief report summarizes the findings and discusses the outcomes 
with a focus 
on Cryogel\textsuperscript{{\tiny{\textregistered}}}, 
while the magnetic properties of 
Pyrogel\textsuperscript{{\tiny{\textregistered}}} are also presented.
After overviewing the samples and methods employed, the low-field 
temperature dependences and isothermal responses of the magnetism are 
presented, anlayzed, and summarized.

\section{Samples and Methods}

The sample-pack received contained nominally (40~cm)$^2$ sheets of five 
materials, specifically the samples (and their thicknesses) were: 
Cryogel\textsuperscript{{\tiny{\textregistered}}}X201 
(5~mm and 10~mm) \cite{CryogelX201},  
Cryogel\textsuperscript{{\tiny{\textregistered}}}Z 
(5~mm and 10~mm) \cite{CryogelZ}, 
and Pyrogel\textsuperscript{{\tiny{\textregistered}}}XTE 
(10~mm) \cite{PyrogelXTE},
whose compositions
are provided in Table~\ref{tab1}. 

\begin{table}[tb!]
\caption{Compositions of Cryogel\textsuperscript{{\tiny{\textregistered}}} and 
Pyrogel\textsuperscript{{\tiny{\textregistered}}} reported 
in this work.}\label{tab1}
\begin{tabular*}{\textwidth}{@{\extracolsep\fill}lcccc}
\toprule%
 &    &  & \multicolumn{1}{@{}c@{}}{Composition (\%)}  
&  \\\cmidrule{3-5}%
 &  
& Cryogel\textsuperscript{{\tiny{\textregistered}}}
& Cryogel\textsuperscript{{\tiny{\textregistered}}}
& Pryogel\textsuperscript{{\tiny{\textregistered}}}\\
Constituent & Formula 
& X201 \cite{CryogelX201}
& Z \cite{CryogelZ}
& XTE \cite{PyrogelXTE}\\
\midrule
Synthetic Amorphous Silca  & SiO$_2$  & $40 - 50$  & $25 - 40$ & $30 - 40$ \\
Methylsilyated Silica  & C$_6$H$_{19}$NSi$_2$ & $10 - 20$  & $10 - 20$  & $10 - 20$ \\
Polyethylene Terephthalate  & C$_{10}$H$_8$O$_4$ & $10 - 20$  & $10 - 20$ &  \\
Fibrous Glass (textile grade)\footnotemark[1]  & SiO$_2$ & $10 - 20$  & $10 - 20$ & $40 - 50$ \\
Magnesium Hydroxide  & Mg(OH)$_2$ & $\;\,0 - \;\,5$  &  $\;\,0 - \;\,5$ &  \\
Aluminum Foil\footnotemark[2]  & Al &  & $\;\,0 - \;\,5$   &  \\
Iron Oxide (Fe(III) oxide)  & Fe$_2$O$_3$ &   &  & $1 - 10$ \\
Aluminum Trihydrate  & AlH$_3$O$_3$ &   &  & $1 - \;\,5$ \\
\botrule
\end{tabular*}
\footnotetext{NOTE: The exact percentages (concentrations) of the 
compositions were withheld as trade secrets.  }
\footnotetext[1]{The Chemical Abstracts Service Registry Number 
 was not given, so our assumption is listed here.}
\footnotetext[2]{Samples were taken in regions away from this foil and its immediate region.}
\end{table}

Samples, with a typical mass, $m$, in the range $20-40$~mg, were extracted 
from the sheets and 
gently pressed directly into a $\sim\,$7~mm long section of the straw 
sample holders, 
which provided a uniform background for the detection 
scheme used by the commercial magnetometer, Quantum Design MPMS XL, capable of 
providing a range of temperature 
$(2~\mathrm{K} \leq T \leq 400~\mathrm{K})$ and magnetic field 
$(-70~\mathrm{kG} \leq B \leq 70~\mathrm{kG})$ conditions.  When studying  
Cryogel\textsuperscript{{\tiny{\textregistered}}}Z, the aerogel-like 
samples were taken from a region 
away from the aluminum foil and its immediate surrounding location.  
The studies of the temperature dependence of the low-field magnetization 
were performed in 
zero-field cooling (ZFC) and field-cooling (FC) modes and were then 
followed by isothermal 
(usually at $T= 5$~K) magnetization measurements while increasing 
($B=0 \rightarrow 70$~kG) 
and then decreasing ($B= 70~\mathrm{kG} \rightarrow - 10$~kG) 
the field to check for hysteresis.  
For this report, the data are expressed in cgs units where a magnetic moment, 
$\mu$, is 1~emu = 1~erg G$^{-1}$ \cite{Blundell2001}, so the following notation 
is employed for the mass magnetization $M = \mu/m$ (emu/g) and 
the mass susceptibility $\chi = M/B$ (emu~g$^{-1}$~G$^{-1}$).  

For inductively-coupled plasma optical-emission spectroscopy (ICP-OES), 
a Pyrogel\textsuperscript{{\tiny{\textregistered}}}XTE or 
Cryogel\textsuperscript{{\tiny{\textregistered}}}X201 sample 
with a mass of 10~mg was dissolved in 20~mL of 1~M KOH solution and 
left overnight to allow the silica to be digested so the
metal ions could dissociate from the fiber. 
The solution was then adjusted to an acidic pH using 1~M HNO$_3$ 
to dissolve any metal oxide and subsequently diluted with 
deionized water to make the final sample solution.  
The ICP-OES study employed a VARIAN VISTA RL simultaneous spectrometer 
(Agilent Technologies, Santa Clara, California, USA) using standard 
addition methods. 

\section{Magnetic Data and Discussion}

\begin{figure}[t]
\centering
\includegraphics[width=0.9\textwidth]{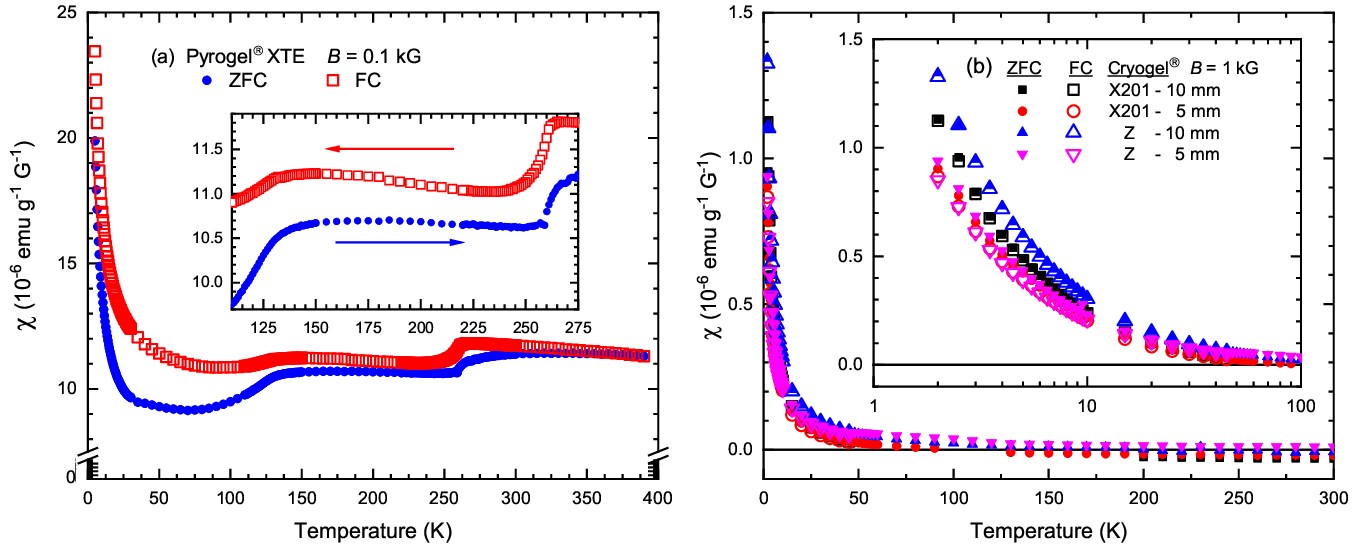}
\caption{The temperature dependent magnetic susceptibilities 
of Pyrogel\textsuperscript{{\tiny{\textregistered}}}XTE 
and the 5~mm and 10~mm thick sheets of 
Cryogel\textsuperscript{{\tiny{\textregistered}}}X210 and 
Cryogel\textsuperscript{{\tiny{\textregistered}}}Z are shown.
(a)~The mass susceptibility of  
Pyrogel\textsuperscript{{\tiny{\textregistered}}}XTE 
measured in $B = 0.1\,\mathrm{kG}$ exhibits a strong 
Curie-like tail at low temperatures, differences between FC and ZFC 
data below a blocking temperature near 130~K, and a sharp shoulder 
at the Morin transition of 260 K, which are assignable to the known 
presence of Fe$_2$O$_3$, see Table~\ref{tab1}.
(b)~The low magnetic field ($B = 1\,\mathrm{kG}$) mass  
susceptibility data for all four samples of 
Cryogel\textsuperscript{{\tiny{\textregistered}}}  
are almost degenerate on a linear 
temperature scale, so the inset shows the resuts on a logarithmic scale 
for $T < 100\, \mathrm{K}$.
In most instances, the ZFC and FC data are the same within measuring 
uncertainty, and the strength of the Curie-like response at low 
temperature is striking.}\label{fig1} 
\end{figure}

In low applied magnetic fields, the initial conjecture was the 
high-temperature magnetic signal 
would be dominated by the diamagnetism of the 
SiO$_2$ matrix \cite{crc105}, and the 
low-temperature response might show evidence of some trace amounts of 
free-spin $S=1/2$ impurities \cite{Sarachik1985,Matsuoka2012} for 
Cryogel\textsuperscript{{\tiny{\textregistered}}},  
whereas signatures reminiscent of the well-studied 
magnetism of  Fe(III)$_2$O$_3$ \cite{Zorbil2002} were  anticipated for 
Pyrogel\textsuperscript{{\tiny{\textregistered}}}.  
Although some of these signatures appeared in the data, 
Fig.~\ref{fig1}, other features were not anticipated.  

More specifically for Pyrogel\textsuperscript{{\tiny{\textregistered}}}XTE, 
Fig.~\ref{fig1}(a), the differences between ZFC and FC traces 
below a blocking temperature of $\approx130$~K and a strong 
Curie-like tail were 
not surprising \cite{Maldonado2017}, while the appearance of the magnetic 
{\emph ``fingerprints''} of a Morin transition at 260~K 
\cite{Morin1950,Suber1998,Kuvaniova2019} were not anticipated.  
Since Fe(III) oxide is listed as an ingredient of 
Pyrogel\textsuperscript{{\tiny{\textregistered}}}, see Table~\ref{tab1}, 
the observed magnetic behavior is consistent with the presence of a 
broad distribution of Fe$_2$O$_3$  nanoparticles with 
diameters in the range of $10\!-\!100$~nm \cite{Suber1998,Kuvaniova2019}.

In contrast to Pyrogel\textsuperscript{{\tiny{\textregistered}}}, 
the high temperature magnetic responses of 
Cryogel\textsuperscript{{\tiny{\textregistered}}} 
are  dominated by the diamagnetic  signal, which is eventually overcome 
by a paramagnetic contribution as  the temperature is lowered, 
Fig.~\ref{fig1}(b). 
Consequently, some gaps in the data sets appear as the signal changes sign 
passing through zero when nearly equal amounts of diamagnetic and 
paramagnetic signals are present. 

\begin{figure}[t!]
\centering
\includegraphics[width=0.9\textwidth]{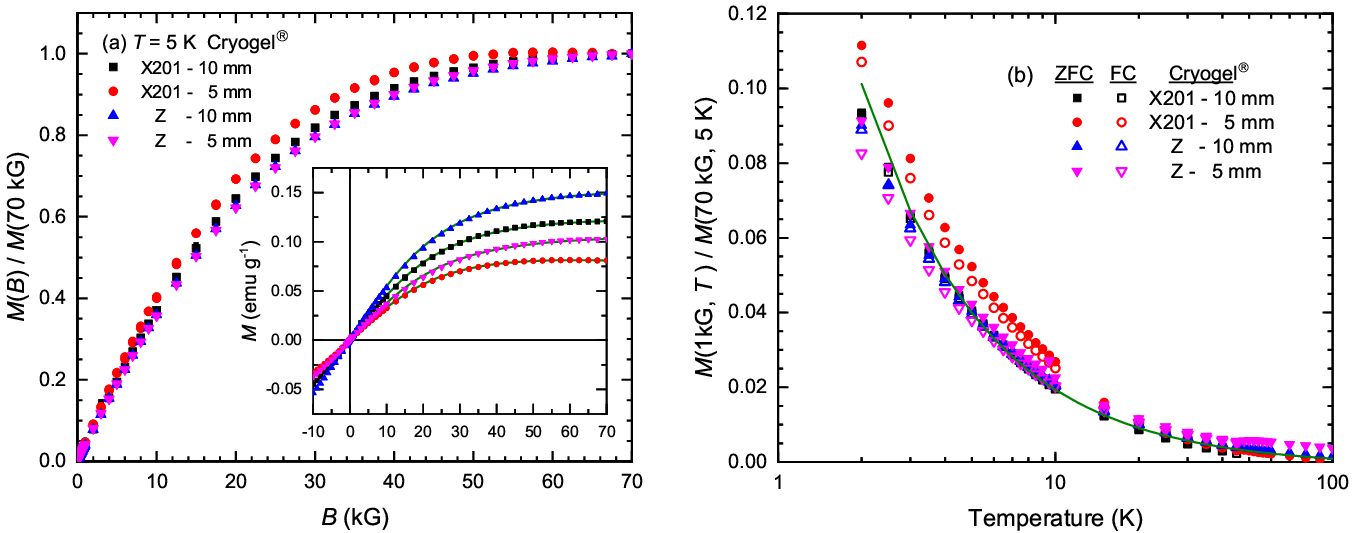}
\caption{(a) The magnetic field dependences of the isothermal 
magnetic moments, $M(B$,~5~K) of 
Cryogel\textsuperscript{{\tiny{\textregistered}}} 
are shown per gram of sample in the inset, and in dimensionless  
form when normalized to their 
$M$(70~kG,~5~K) values.
(b)  The data from the Fig.~\ref{fig1}(b) 
inset are replotted in dimensionless form. 
The green lines for the inset of (a) and for (b) represent the results 
of the model, see text and parameters in Table~\ref{tab2}.  
For clarity, the X201 - 10 mm result is shown in (b).}\label{fig2}
\end{figure}

A striking aspect of the magnetism of the 
Cryogel\textsuperscript{{\tiny{\textregistered}}} samples is  
the strong strength of the Curie-like tail below nominally 50~K, 
Fig.~\ref{fig1}(b).  Since all of the data appear to be 
almost degenerate on a linear temperature scale, the 
inset shows an expanded view of the data on a logarithmic 
scale where essentially any differences between the ZFC and 
FC data are within the experimental resolution.  
Subtle differences between the four samples are noticable and 
were also  detected in the low-temperature isothermal 
magnetization studies, as shown in the inset of Fig.~\ref{fig2}(a).  
In fact, these results capture a major issue 
 of  establishing the mass of the sample being studied.  
 Extracting samples from the sheets involved 
 cutting the materials in various ways, and 
 all methods  led to the  samples shedding 
 small shards of their contents when being 
 handled, with some material 
 detected in the tape-end-cap at the bottom 
 of the straw used for the magnetometry 
 studies.   Consequently, these data were normalized to their 
 $M$(70~kG,~5~K)  
 values,  see Table~\ref{tab2},
 and with the exception of the results for  
 Cryogel\textsuperscript{{\tiny{\textregistered}}}X201 -- 5 mm, 
 a universal trend is established within 
 the experimental uncertainty of the 
 magnetic signals, Fig.~\ref{fig2}(a). 
 To further clarify this issue, the data 
 in the inset of  Fig.~\ref{fig1}(b) 
 were normalized to their 
 $M$(70~kG,~5~K) 
 values, Fig.~\ref{fig2}(b), which 
 also provides a sense of the 
 magnetic response  
 independent of the mass 
 of the sample.

\begin{figure}[th!]
\centering
\includegraphics[width=0.9\textwidth]{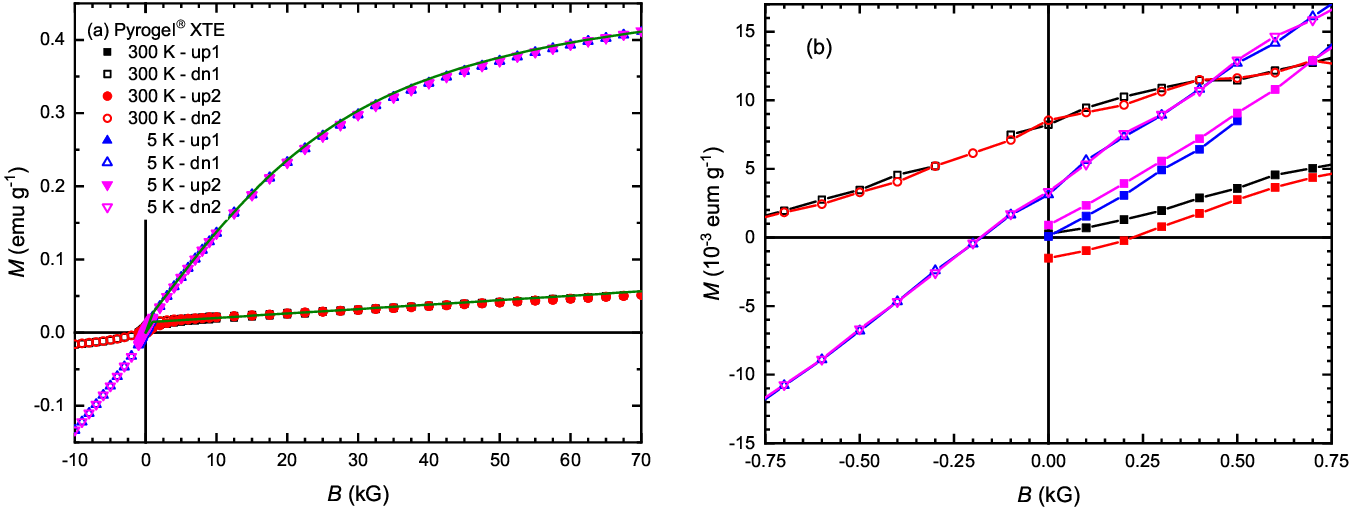}
\caption{The magnetic field dependences of the isothermal magnetic moments, 
$M(B$, 5~K and 300~K) of 
Pyrogerl\textsuperscript{{\tiny{\textregistered}}}XTE 
are shown in (a) for up and down (dn) sweeps as described in the 
text and with the modeling results shown by solids lines, 
and in (b) as an expanded view 
in the region near the origin and 
where the lines connect the data
points.}\label{fig3}
\end{figure}

The isothermal magnetization of 
Pyrogel\textsuperscript{{\tiny{\textregistered}}}XTE 
is shown in Fig.~\ref{fig3}, where 
up1/dn1 refers to the field 
sweeping up/down.  At 300~K, the first 
measurement was made immediately after 
inserting the sample into the magnetometer, 
and the second run was made after the sample 
was ``degassed" while measuring in 0.1~kG 
from 300~K to 390~K over a period of 4~h 
before cooling the sample back to 300~K.  
During the heating cycle, the magnetic 
signal subtly decreased for the first 90 min and 
was then independent of the conditions.  
Although the initial parts of each $M(B$,~300~K) run are 
slightly history dependent, the return to 
zero field conditions indicates  no 
substantial impact of the magnetic response due to the heating cycle, see 
Fig.~\ref{fig3}(b). 
Two different studies were also 
conducted at 5~K to check reproducibility 
and history dependences of the sample, and the overall magnetic 
response was determined to be robust, Fig.~\ref{fig3}.

\begin{table}[bh!]
\caption{Details of samples measured and parameters providing 
phenomenological approximations of the magnetic responses.}\label{tab2}
\begin{tabular*}{\textwidth}{@{\extracolsep\fill}lcccc}
\toprule%
 &  File
& mass
& $M(70\,\mathrm{kG},5\,\mathrm{K})$
& $\chi_{\mathrm{o}}$\\
Sample &  ID 
& (mg)
& (emu g$^{-1}$)
& $(10^{-6}\,\mathrm{emu}\,\mathrm{g}^{-1}\,\mathrm{G}^{-1})$\\
\midrule
Cryogel\textsuperscript{{\tiny{\textregistered}}}X201 - 10 mm  & 230206  & 28.96  & 0.12095 & $-1.0$ \\
Cryogel\textsuperscript{{\tiny{\textregistered}}}X201 - 5 mm  & 230126 & 27.18  & 0.08105  & $-2.5$ \\
Cryogel\textsuperscript{{\tiny{\textregistered}}}Z - 10 mm  & 230201 & 29.77  & 0.14906 & $-0.5$ \\
Cryogel\textsuperscript{{\tiny{\textregistered}}}Z - 5 mm  & 230214 & 35.72  & 0.10286 & $-0.5$ \\
Pryogel\textsuperscript{{\tiny{\textregistered}}}XTE - S1\footnotemark[1]  & 220623 & 22.02  &  NA &  $---$\\
Pryogel\textsuperscript{{\tiny{\textregistered}}}XTE - S2\footnotemark[1]  & 220701 & 24.23  &  0.41118 &  $+1.0$\\
\botrule
\end{tabular*}
\footnotetext[1]{Two different samples, S1 for Fig.~\ref{fig1}(a) and 
S2 for Fig.~\ref{fig3}.}
\end{table}

\newpage

\begin{figure}[bh!]
\centering
\includegraphics[width=0.45\textwidth]{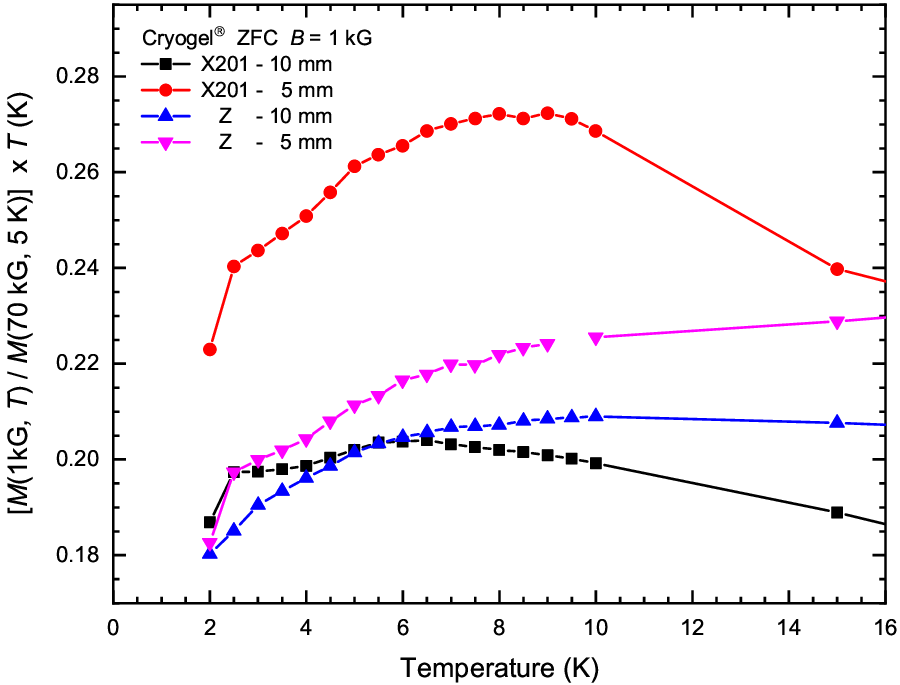}
\caption{The normalized, $B=1$~kG data for 
Cryogel\textsuperscript{{\tiny{\textregistered}}} 
shown in Fig.~\ref{fig2}(b) are multiplied by 
temperature and are replotted on a linear scale below 15~K. 
The solid lines connect neighboring data points 
and the inferences are discussed in the text.}\label{fig4}
\end{figure}

Lastly, since the low-field plots 
of Fig.~\ref{fig1}(b) and \ref{fig2}(b) 
do not elucidate subtle 
trends that may exist in the region of 
the strong Curie-like tail, the 
mass-independent normalized data 
of Fig.~\ref{fig2}(b) were used 
to generate the effective 
$\chi \times T$ versus $T$ plot of Fig.~\ref{fig4}, 
where additional fidelity is revealed. 
For the instrument being used, the cooling 
mechanism employed switches below 4.5~K, 
resulting in additional time to cool samples 
to the minimum temperature.  Consequently, as 
is the case in these data sets, the $T=2$~K data point 
possesses evidence of not being in thermal 
equilibrium before the measurement was performed.  
From this viewpoint, the $T \lesssim 10$~K 
data, acquired over a period of 1~h after 
stabilizing at $T=2$~K, may have been 
acquired when the sample was not 
in thermal equilibrium with the thermometer 
of the instrument.  However, the 
isothermal magnetization 
studies performed at 5~K (chosen to avoid 
concerns about thermal equilibrium) 
take nominally 5~h to complete 
but do not show any hysteresis that would 
arise from non-equilibrium conditions.

Ensemble the results motivated a check of the level 
of magnetic species that might be present in the 
samples, an ICP-OES study was performed. 
With no other metal being detected, Fe was detected 
at 0.26\%w/w for 
Pyrogel\textsuperscript{{\tiny{\textregistered}}}XTE 
and at 0.09\%w/w for 
Cryogel\textsuperscript{{\tiny{\textregistered}}}X201. 
These values are considered as lower bounds due to the 
incomplete dissociation of the silica in the samples.  
The Fe content levels were not explored in greater 
detail since this thrust was beyond the scope of this work.

\section{Phenomenological Model}

At the start of this work, the goal was to characterize the 
magnetism of 
Cryogel\textsuperscript{{\tiny{\textregistered}}} 
arising from conjectured trace amounts of non-interacting 
species, with total angular momentum $J$ and $g$-factor values, 
which might be described the 
Brillouin function, $\mathcal{B}_J(J,g,B,T)$ \cite{Blundell2001}. 
Specifically for a sample of mass $m$ and with $N$ entities, the 
predicted mass magnetization can be written as
\begin{equation}\label{eq1}
M(B,T) \; = \; \frac{N}{m}\,J\,g\,\mu_{\mathrm{B}}\, \mathcal{B}_J (g,J,B,T)\;\;\;,
\end{equation}
where $\mu_{\mathrm{B}}$ is the Bohr magneton.  Given the difficulty in 
establishing the mass of the samples, each side Eq.~\ref{eq1} can 
be normalized by the measured $M$(70~kG,~5~K) values for each sample to yield
\begin{equation}\label{eq2}
\frac{M(B,T)}{M(70\,\mathrm{kG},\,5\,\mathrm{K})}\; =\;
\frac{\mathcal{B}_J (g,J,B,T)}{\mathcal{B}_J (g,J,70\,\mathrm{kG},\,5\,\mathrm{K})}
\;+\; \frac{\chi_{\mathrm{o}}\;B}{\mathcal{B}_J (g,J,70\,\mathrm{kG},\,5\,\mathrm{K})}
\;\;\;,
\end{equation}
where a temperature independent term is added to accommodate the 
diamagnetism from the silca.
Equation~\ref{eq2} provides motivation for generating Fig.~\ref{fig2} and 
also yields the green lines when using the values listed in Table~\ref{tab2} 
with $J=5/2$ and $g=2.03$ \cite{Goldfard1994,Jahagirdar2013}.

With respect to the results for 
Pyrogel\textsuperscript{{\tiny{\textregistered}}}XTE, 
there is no basis for a non-interacting spin model to be valid. 
Nonetheless, Eq.~\ref{eq2} provides the green lines shown in 
Fig.~\ref{fig3}(a) when 
using the values listed in Table~\ref{tab2} with $J=5/2$ and 
$g=2$ \cite{Goldfard1994,Jahagirdar2013}, albeit with the accommodation of 
a constant 0.014~emu~g$^{-1}$ remanent magnetization, Fig.~\ref{fig3}(b).

\section{Summary}

By providing a survey of the static magnetic properties of 
Cryogel\textsuperscript{{\tiny{\textregistered}}} and 
Pyrogel\textsuperscript{{\tiny{\textregistered}}}XTE 
at low temperatures and in high magnetic fields, this brief report 
fills a void in the literature about these properties which need to be 
known before deployment in some potential applications.    
A phenomenological model provides reasonable estimates 
for the magnetism observed, but of course,
the specific outcomes are likely to be fabrication batch dependent on an 
industrial scale, and this reason may explain the X201~--~5~mm 
results being different than the responses detected from the other 
Cryogel\textsuperscript{{\tiny{\textregistered}}} materials.

\newpage

\bmhead{Acknowledgements}
The National Science Foundation (NSF) Research Experiences for 
Undergraduates (REU) funding provided 
support for the participation of: QLW and AJS 
(Fall 2021 and Spring 2022) and CLB (Spring 2022) via 
DMR-1708410; GTH and RJR (Summer 2022) and CLB (Fall 2022) via 
MagLab REU Program DMR-1644779; 
CLB (Summer 2022) via UF Condensed Matter and Applied Materials 
REU Program DMR-1852138, 
which also provided professional development and social networking 
activities to GTH and RJR (Summer 2022); and CLB 
(Spring 2023, Fall 2023, and Spring 2024) via MagLab DMR-2128556.  
The University of Florida University Scholars Program (USP) and 
Center for Undergraduate Research (CUR) provided additional support to 
CLB (Fall 2023 and Spring 2024). 
Aspects of this work also used facilities and personnel supported by the 
National High Magnetic Field Laboratory (NHMFL or MagLab), via 
NSF Cooperative Agreement DMR-1644779 
and DMR-2128556, and the State of Florida.  
Patience on the part of all stakeholders is gratefully 
acknowledged as this work was 
initiated during the pandemic.

\bibliography{Cryogel-2025}

\end{document}